\documentstyle[11pt]{article}
\begin{document}
 
\begin{titlepage}

\begin{center}
 
\vspace{.5 true  cm}
 
\Large
{\bf The magnetism of the
$t-t'$ Hubbard model } \\
\vspace{6pt}

\vspace{1.5 true  cm}
 
\large
{{\bf B. Valenzuela\S \dag \footnote{e-mail: belen@jambo.icmm.csic.es},
M.A.H. Vozmediano\dag\footnote{e-mail: vozmediano@pinar1.csic.es}, and
F. Guinea\S \footnote{e-mail: paco.guinea@uam.es}}}

\vspace{0.5 true  cm}
 
\normalsize
 
\S{\em Instituto de Ciencia de Materiales, CSIC} \\
{\em Cantoblanco, 28049 Madrid, Spain.} \\
 
\dag{\em Departamento de Matem\'aticas, Universidad Carlos
III de Madrid} \\
{\em Avda. de la Universidad 30, 28913 Legan\'es (Madrid), Spain.}
 
\vspace{.5 true  cm}

\end{center}
 
\begin{abstract}

The magnetic properties of the $t-t'$ Hubbard model in the two dimensional
square lattice are studied within an unrestricted Hartree
Fock approximation in real space. The interplay between
antiferromagnetism, ferromagnetism, phase separation and inhomogeneous
magnetic textures is studied. It is shown that, at sufficiently
large values of $t'/t$, a rich fenomenology is to be expected
between the antiferromagnetic phase at half filling and the
ferromagnetic phase at lower fillings.

\vspace{3 true cm}

\end{abstract}

\end{titlepage}
\newpage

\section{Introduction}

The single band  Hubbard model is widely accepted 
as the simplest starting point 
of a microscopic description of correlated electron systems \cite{and}.
More recently it has been realized that inclusion of a $t'$ coupling \cite{ttp}
can fit the phenomenology of some cuprates. Recent photoemission
experiments  seem to provide Fermi surfaces  compatible with 
the dispersion relation of the model at  moderate values of $t'$ and densities
\cite{arpes}, while it has been argued that it can fit some
features of ruthenium compounds \cite{rut} at higher values of $t'$ and 
the doping.\\

The study of inhomogeneous charge and spin phases in the Hubbard model has been a 
subject of interest since the discovery of the high-$T_c$ compounds  as
it was seen that they have a very inhomogeneous electronic structure
at least in the underdoped regime. However most of the work was done 
before the importance of $t'$ was realized \cite{su,pr,il,paco1}.
A sufficiently large value of the ratio $t'/t$, compatible with
the values suggested for the cuprates ($t'/t \sim -0.3$),
leads to a significant change in the magnetic properties of the model,
as a ferromagnetic phase appears at low doping. This phase
has been found by numerical and analytical 
methods \cite{ferro1,ferro3,phases}, and
it is a very robust feature of the model.\\

The purpose of this work is to study the influence of $t'$ on the 
magnetism of the Hubbard model at moderate values of U and density. 
We will make use of an unrestricted 
Hartree-Fock approach in real space what allows us to visualize the 
charge and spin configurations. We believe that the method is 
well suited for the present purposes as: i) it gives a reasonable
description of the N\`eel state, with a consistent description of
the charge gap and spin waves, when supplemented with the RPA.
ii) It becomes exact if the ground state of the model is
a fully polarized ferromagnet, as, in this case, the interaction
plays no role. A fully polarized ground state is, indeed, compatible
with the available Montecarlo\cite{ferro1} and t-matrix 
calculations\cite{ferro3}. iii) It is a variational technique, and it should
give a reasonable approximation to the ground state energy.
This is the only ingredient required in analyzing the issue
of phase separation. iv) It describes the doped antiferromagnet,
for $t'=0$ as a dilute gas of spin polarons. The properties of
such a system are consistent with other numerical calculations
of the same model \cite{letal98,letal99}.\\

On the other hand, the method used here does not allow us to treat
possible superconducting instabilities of the model, which have been
shown to be present, at least in weak coupling approaches \cite{phases}.
The study of these phases requires extensions of the present
approach, and will be reported elsewhere.\\

The main new feature introduced by a finite $t'$, using simple
concepts in condensed matter physics, is the destruction the perfect
nesting of the Fermi surface at half filling, and the existence
of a second interesting filling factor, at which the 
Fermi surface includes the saddle points in the dispersion
relations. At this filling, the density of states at the
Fermi level becomes infinite, and the metallic phase
becomes unstable, even for infinitesimal values of the
interaction. For sufficiently large values of $t'/t$, the
leading instability at this filling is towards a ferromagnetic 
state.\\

In the following section, we present the model and the method.
Then, we discuss the results. As the system shows a rich variety
of behaviors, we have classified the different regimes
into an antiferromagnetic region, dominated by short range
antiferromagnetic correlations, a ferromagnetic one,  
and an intermediate situation, where the method suggest
the existence of phase separation. The last section presents
the main conclusions of our work.

\section{The model and the method}

The t-t' Hubbard model is defined in the two dimensional
squared lattice by the hamiltonian

\begin{equation}
H=-t\sum_{<i,j>,s} c^+_{i,s}
c_{j,s}
\;-t'\;\sum_{<<i,j>>,s} c^+_{i,s} c_{j,s} 
\;+\;U\sum_i n_{i,\uparrow} n_{i,\downarrow}
 \;\;\;,
\label{ham}
\end{equation}
with the dispersion relation 
$$
\varepsilon({\bf k}) = 2t\;[ \cos(k_x a)+\cos(k_y a)]
+4t'\cos(k_x a)\cos(k_y a)
\;\;\;.
$$

We have adopted the convention widely used to describe the phenomenology 
of some hole-doped cuprates \cite{ttp}, $t>0$, $t'< 0 \;,\; 2t'/t <\; 1\;$. 
With this choice of parameters the bandwidth is $W=8t$ and
the Van Hove singularity 
is approached by  doping the half-filled
system with holes. Throughout this study we will fix the value of $t=1$ 
so that energies will be expressed in units of $t$. Unless otherwise stated, 
we will work in a $12\times 12$ lattice with periodic boundary conditions. 
We have choosen the $12\times 12$ 
lattice because it is the minimal size for which
finite size effects are almost irrelevant
\cite{pr,pilar,capone}.  \\
 
The unrestricted Hartree-Fock approximation 
minimizes the expectation value of
the  hamiltonian (\ref{ham}) in the space of Slater determinants. These are 
ground states of a single particle many-body system in a potential 
defined by the electron occupancy of each site. This potential is
determined selfconsistently
$$
H=-\sum_{i,j,s} t_{ij} c^\dag_{i,s}c_{j,s} 
-\sum_{i,s,s'}\frac{U}{2} {\vec m}_i c^\dag_{i,s} {\vec \sigma}_{s,s'}  c_{i,s'}
+\sum_{i}\frac{U}{2} q_i(n_{i\uparrow}+n_{i\downarrow})  +
{\displaystyle c.c.}\;\;,
$$
(where $t_{i,j}$ denotes next, t,  and next-nearest, t', neighbors), 
and the self-consistency conditions are
$${\vec m_i}=\sum_{s,s'}<c^\dag_{i,s} {\vec \sigma}_{s,s'}  c_{i,s'}> \;\;\;,\;\;\;
q_i=< n_{i\uparrow}+n_{i\downarrow}-1> \;,$$
where ${\vec \sigma}$ are the Pauli matrices. \\

We have established a very restrictive criterium
for the convergence of a solution. The iteration ends when
the effective potential of the hamiltonian and the one deduced from the
solution are equal up to $E\;<\;10^{-7}$.
When different configurations converge for a given value of the parameters,
their relative stability is found by comparing their total energies.

\section{The results}  

The results in this work are summarized in fig.1
which represents the energy of the 
ground state configurations versus doping from $x=0$ to $x=0.34$ 
(where x is the rate of total number of electrons over 
the total number of sites) 
for the representative 
values $t'=0.3$ and $U=8$. As, in most cases, a variety
of selfconsistent solutions can be found, we  have tried to avoid an inital
bias by starting  with random spin and charge configurations. Once
the system has evolved to a stable final configuration, this has been used 
as initial condition for the nearby dopings. Hence, most of the 
configurations discussed in the text are robust in the sense that they 
have not been forced by a choice of initial 
conditions and hence are stable under
small changes of the initial values. Exceptions
are the diagonal  commensurate domain walls 
and the stripes. These configurations were set 
as initial conditions and found to be self-consistent. Even if
there are many possible solutions,
the system cannot be ^^ ^^ forced " to converge to a given solution
by appropiately choosing the inital conditions.
In particular homogeneous solutions, such as a pure AF solution, do not 
converge near half filling, as will be discussed later. Fig. 2 and fig. 3 
show a comparison of the energies of different configurations converging 
in the same range of dopings. Fig. 1 shows only minimal energy 
configurations. Once a configuration converges, we have checked its 
stability under changes in $U$ and $t'$.\\

The most remarkable feature of fig. 1 is the smooth transition 
from insulating antiferromagnetism to 
metallic ferromagnetism. The  antiferromagnetic 
region extends in a range of hole
doping from 
$x=0$ to $x=0.125$ and the ferromagnetic region from $x=0.125$ to $x=0.34$. 
In the antiferromagnetic region the 
predominant configurations are fully polarized antiferromagnetism (AF), 
polarons (POL), 
diagonal commensurate domain-walls(dcDW), and noncollinear solutions ($S_x$). 
In the ferromagnetic region the phases are ferromagnetic domains (fm DOM), 
ferromagnetic non collinear solutions 
(fm SDW) and the fully polarized state or Nagaoka configuration (Ng).\\

Most of the AF configurations are known as solutions of the Hubbard model
with $t'=0$ \cite{su,pr,il,paco1,zaan,Sc90}. We will here comment 
on the changes induced by $t'$. The FM configurations are totally new 
and due to the presence of $t'$, as well as 
the zone of coexistence of both magnetic orderings.
In addition, we have also analyzed in 
detail  some striped configurations, due to their
possible experimental relevance.\\

In what follows we analyze the antiferromagnetic and ferromagnetic regions 
and discuss the possibility of phase separation.

\subsection{The antiferromagnetic region} 

The study of the motion of a few holes in an antiferromagnetic background
has been one of the main subjects in the literature related to the cuprates 
as these are doped AF insulators. The region of the Hubbard model
at and close to half filling is also the area where the 
metal-insulator transition \cite{imada} occurs, and where the well-established
spin polarons or spin bags \cite{bag} 
coexist with domain walls and, possibly, striped
configurations. 
The diagonal hopping $t'$ has a strong influence over this region as 
it destroys the perfect nesting 
of the Hubbard model at half filling and the particle-hole symmetry 
which leads to AFM order in weak coupling approaches \cite{lin,vollhardt}.\\

The AF region is formed by fully polarized antiferromagnetism, 
polarons, diagonal commensurante 
domain-walls and non collinear solutions. We also have found stripes 
as excited 
states. The configurations and the density of states are shown in 
fig. 4, fig. 5, fig. 6 and fig.7. We will give a brief 
discussion of these configurations. 

\vspace{0.5 true cm}
 
{\bf Antiferromagnetism}

\vspace{0.3 true cm}

For the reference values of $U=8$ and $t'=0.3$ fully polarized 
antiferromagnetism (AF) is the lowest energy configuration  
only at half filling. For the range of dopings  
$0.007\leq x\leq 0.027$, ($1\leq h\leq 4$, where h denotes the 
number of holes), AF converges but  
POL are energetically 
more favourable. Above four holes a purely AF initial configuration 
evolves to polarons. \\

AF is the minimal energy configuration for lower values
of U in a wider range of dopings. 
For example for $U=4$ and $t'=0.3$ AF is the lowest energy 
configuration in the range 
$0\leq x\leq 0.03$. This result is almost insensitive to changes in $t'$.  \\

We can conclude then that in the presence of $t'$,
the homogeneous fully polarized antiferromagnetic configuration (N\`eel state)
is not the dominant solution near half filling. 
This result is to be contrasted
with what happens with electron doping where AF dominates a larger region
of the doping space. The reason for this asymmetry will become clear in the
discussion of the polaronic configuration following. For $t'\neq 0$
inhomogeneous  solutions are clearly energetically more favourable. 

\vspace{0.5 true cm}

{\bf Polarons}

\vspace{0.3 true cm}
Magnetic polarons have been discussed at length in the literature 
\cite{bag,paco1,seib}. For $t'=0$ the magnetization points along the 
same direction everywhere in the cluster and the extra charge is localized 
in regions that can be of either cigar or diamond shape. These regions 
defined a core where the magnetization is reduced. \\

In the present case, $t' \neq 0$, this picture changes substantially.
The two Hubbard bands in the N\`eel state are no longer equivalent,
with bandwidths given, approximately, by $8 | t' | \pm 4 t^2 / U$.
Polarons are found at the edges of the narrower band at all values of t'.
 This
situation corresponds to hole doping for our choice of sign
of $t'$ ( $t' / t < 0$). The doping of the wider
electron band leads usually
to stable homogeneous metallic AF solutions,
where the extra carriers are delocalized throughout the lattice.
We have found polarons in the electron region
only when $U$ is big ($U > 6$) and t' small
($t' < 0.15$). On the other hand, the localization of the polarons
induced by hole
doping increases with increasing $|t' / t |$. This reflects the
fact that these polarons are derived from a narrower Hubbard band.
Qualitatively, this fact can be understood
in terms of the asymmetric tendency of the system towards
phase separation when $t' \neq 0$ \cite{Getal99}.
The polarons also can be understood as an incipient form of phase
separation, as the core shows strong ferromagnetic correlations.\\

Polarons converge in a wide region of the phase diagram, 
coexisting with AF and dcDW as shown in fig. 2. They have lowest energy 
in the doping range  
 $0.007\leq x\leq 0.035$ ($1\leq h\leq 5$). They do not 
converge in the range $0.076\leq x\leq 0.097$ ($11\leq h \leq 13$) where 
the noncollinear solution has lower energy. 
This is also different from 
the situation with $t'=0$ where polarons converge and have lower 
energy in the full range of dopings $2\leq h\leq 30$ \cite{paco1}.  
The DOS of polarons is shown if fig. 5b for five holes. 
Fig. 5a shows the reference AF state at half filling. In the DOS for 
polarons the localized states appear in the antiferromagnetic gap. 
As doping increases, they form a mid gap subband but the shape of 
antiferromagnetic spectrum 
is still clearly seen (see fig. 5c). \\

\vspace{0.5 true cm}

{\bf Diagonal commensurate domain-walls}

\vspace{0.3 true cm}

The dcDW configuration is formed by polarons arranged in the diagonal 
direction creating an almost one dimensional charged wall that forms
a ferromagnetic domain (see
fig. 4b). We stress ^^ ^^ commensurate "  because they do not separate 
different AF domains as
stripes do. 
The density of states of these solutions differ from
that of usual domain walls in that the one dimensional
band where the holes are located is wider, and the Fermi level
lies inside it. We do not find a size independent one particle gap
in these solutions, unlike for conventional domain walls.
Within the numerical precision of our calculations, these
structures are metallic, while antiphase domain walls are
insulating.\\ 
 
The dcDW are the predominant configuration in the 
AF region. They converge in the range of dopings $0.014\leq x\leq 0.118$ 
($2\leq h\leq 17$) 
and are the minimal energy configuration in most of the doping range as can be 
seen in fig. 2. These configurations resemble an array of
the polarons discussed earlier, along the (1,1) direction.
Thus, we can say that individual polarons have a tendency towards
aligning themselves along the diagonals.
This may be due to the 
fact that $t'$ favors the hopping along these directions.  
This tendency  can be also seen in the density of states.
We have also
checked that a lower value of $t'$ ( also for $U=8$)
these configurations are less favored.
In particular for $U=8$, dcDW are not 
formed at low $t'$ while they do form at $t'= 0.3$.  
Moreover we have tried 
vertical domain walls and they are not formed for $U=8$ and $t'=0.3$ 
but they do with $t'$ lower ($t'=0.1$). Summarizing, we have checked that $t'$ 
favours dcDW against POL and disfavours vertical DW. 
This behaviour is also found with the striped configurations 
(see below).

\vspace{0.5 true cm} 

{\bf Non collinear solution $S_x$}

\vspace{0.3 true cm}

The structure denoted by $S_x$ 
in the phase diagram consists of a special configuration 
with noncollinear spin $S_z$. It appears in the range of dopings 
$0.076\leq x\leq 0.097$, 
($11\leq h\leq 13$), in competition with
dcDW and has lower energy. We have checked that this structure is 
never seen in the absence of $t'$ and is also found at $U=4$ and $t'=0.3$.\\

The configuration is shown if fig. 4c. We see that there are polarons but there
is a contribution of the spin x at some random sites. 
The convergence in this 
configuration is very slow. It is interesting to point out 
that we obtain this configuration when using conventional  polarons as 
the initial condition.\\

We do not have a complete understanding of why this configuration is preferred 
in this region although, it is interesting to note that it happens for the 
commensurate value of twelve holes 
(in our $12 \times 12 $ lattice) and the two neighboring values $h=11$ and 
$h=13$.\\

Its density of states shown in fig. 5d is very similar to the polaronic DOS.

\vspace{0.5 true cm}

{\bf Stripes}

\vspace{0.3 true cm}

The striped configurations are 
similar to the domain-walls but the one dimensional
arrange of charge separates  antiferromagnetic domains with a 
phase shift of  $\pi$.  Two
typical striped configurations  are shown in 
fig. 6a and fig. 6b. Recently the stripes have attracted 
a lot of interest as the half filled vertical 
stripes (one hole every second site) are found in cuprates while diagonal
stripes are found
in  nickelates \cite{strexperim}.\\

We have obtained stripes as higher energy configurations
and we have not found half 
filled vertical stripes in agreement  
with other works using mean field approximations 
for $U=8$ \cite{su,pr,il,paco1,zaan}.
It is known that the addition 
of a long range Coulomb interaction could stabilize the vertical stripes 
as ground 
states for large $U$ \cite{zaanen}  
and applying a slave-boson version of the Gutzwiller 
approach the half filled vertical stripes can be ground states depending 
on paramenters \cite{seibold}.\\

We have studied the filled vertical stripe and the diagonal stripe obtained
at values of the doping
commensurate with the lattice.
Our main interest is the role played by $t'$ in these configurations. 
We have seen that 
$t'$ has a strong influence on them: $t'$ reduces
significantly the basin of attraction of the vertical stripes, what agrees 
with recent calculations  in the $tt'-J$ model \cite{tt'-J},
while it favors diagonal stripes. The evolution of the vertical stripe 
with $t'$ 
can be seen in fig. 6b and fig. 6c for the values $U=4$, $t'=0$ and $t'=0.2$ 
. These stripes do not converge for higher values of $t'$.
We have instead found that diagonal stripes are favored by $t'$ much as
the dcDW were. \\

The density of states of the stripes is very similar to that of polarons. 
We can conclude that they are insulating states (see fig. 
7) unlike the similar commensurate domain walls where a
more metallic character can be appreciated.

\subsection{The ferromagnetic region}

The existence of metallic ferromagnetism in the Hubbard model 
remains one of the most 
controversial issues in the subject \cite{ferro4}. 
Large areas of ferromagnetism in the doping 
parameter were found in the earliest works on 
the $t-t'$ Hubbard model within mean field 
approximation \cite{lin} and were often assumed to be 
an artifact of the approximation which 
would be destroyed by quantum corrections. 
There are two main regions where ferromagnetism 
is likely to be the dominant configuration. 
One is the region close to half filling, in particular at one hole doping where
the Nagaoka theorem 
ensures a fully polarized ferromagnetic state 
in a bipartite lattice at $U=\infty$. The other
is the region around the Van Hove fillings 
where there is  a very flat lower band and
where ferromagnetism was found for large values of $t'$ close to $t'= 0.5$
with quantum Monte Carlo techniques \cite{ferro1}
and in the T-matrix approximation \cite{ferro3}. 
FM  is also found to be the dominant
instability for small $U$ and large $t'$ in 
analytical calculations based on the renormalization 
group \cite{phases}, and at intermediate 
values of U with a mixture of analytical and mean field 
calculations \cite{japon}. 
Finally, there is a controversy on whether Nagaoka ferromagnetism 
is stabilized at the bottom of the band  $\rho\rightarrow 0$ \cite{botton}.\\

Most of the previous calculations relay on the study of the divergences of
the  magnetic susceptibility
pointing to either a symmetry breaking ground state or to the formation of spin 
density waves  as low energy excitations of the  system. In many cases
it is not possible in this  type of analyses to
discern on the precise nature of the magnetic phases, and,
in particular, whether they correspond
to fully polarized  states (long range order)  or  
to inhomogeneous configurations with
an average magnetization. 
A complete study of the magnetic transitions as a function
of the electronic density is also a difficult issue. \\

We have studied the stability of ferromagnetic configurations in the full
range of dopings discussed previously. Two main issues can be addressed
within the method of the present paper. One is the existence 
of the fully polarized ferromagnetic state (Nagaoka state),
and its stability not only towards the state with one spin flip, 
but against any weakly polarized or paramagnetic configuration.
The other is the specific symmetry of the partially polarized
ferromagnetic  configurations.\\

In the region close to half filling, our results indicate 
that the Nagaoka theorem does probably hold in the
presence of $t'$ (which spoils the bipartite character of the lattice) 
since the
Nagaoka state appears when doping with one hole at such
large values of $U$ as to make the kinetic 
term quite irrelevant. We found Nagaoka FM
at values of $U$ such as $U=128$. 
No FM configurations are found doping with two holes even at $U=128$.\\

The region of low to intermediate  electron density has been analyzed for 
various values of $U$ and $t'$. 
This region includes dopings close to the Van Hove singularity
where FM should be enhanced due to the large degeneracy of states in the 
lower band. The position
of the Van Hove singularity 
for a given value of $U$ and $t'$ can be read off from the undoped DOS;
it has been determined in \cite{vh}.
Our results are the following: \\

Nagaoka FM is not found for $t'= 0.1$ at any filling for $U\leq 8$. For
$t'= 0.3$,  two types of 
FM configurations are the most stable in the range of dopings shown in
fig. 1. Ferromagnetic spin density waves (fm SDW) despicted in
fig. 8b dominate the phase diagram at 
densities close to the AFM transition $0.146 \leq x\leq 0.194$
($21\leq h\leq 28$), and in $x\geq 0.264$ ($h\geq 37$). 
In the region in between, ferromagnetic domains
(fm DOM) as the one shown in fig. 8c are the most stable. 
Both types of configurations show a strong charge segregation and are clearly
metallic. The excitation spectrum of these configurations can be seen 
in fig. 9.\\

Fully polarized FM metallic states 
(Nagaoka) shown in fig. 8a, are found 
at all values of $h$ corresponding to closed shell configurations
from a critical value  
$h_c (U)$ depending on $t'$ till the bottom of the band. They are
shown as vertical solid lines in fig. 1. They are also metallic with a
higher DOS at the Fermi level than the partially polarized configurations. 
 Larger values of $t'$ or $U$ push down the critical $h$
in agreement with previous works \cite{lin,ferro3}.
Some values of $h_c (U)$ are, for $t'=0.3$, 
$h_c(6)=37 , h_c(8)=29 , h_c(10)=21\;.$
For $t'=0.4 , h_c(8)=25\;.$ As mentioned before, 
no FM is found for $t'=0.1$. \\

The former results show a large region of 
ferromagnetic configurations  whose upper
boundary coincides with previous 
estimations \cite{ferro3} but which extends to the
botton of the band. 
We have found paramagnetic configurations to converge in the bottom
of the band but their energies are higher than the ferromagnetic ones. 
Comparing our solutions with the corresponding results obtained with
the same method in the case $t'=0$ \cite{paco1}, 
we find that the inclusion of $t'$ favors 
ferromagnetism for intermediate to large dopings.

\subsection{Phase separation}

Although the issue of phase separation (PS) in the Hubbard model is  quite
old  \cite{visscher}, it has become the object of 
very active research work \cite{allps}
following the experimental
observation of charge segregation 
in some cuprates \cite{psexp}. Despite the effort,
the theoretical situation is quite controversial,
although recent calculations rule out PS in the 2D
Hubbard model \cite{nops,capone}. It seems to occur
above some values of $J$ \cite{someps,Cetal98,CBS98}, although other
work suggests that it is likely
for all values of $J$ in the $t-J$ model \cite{tj}. 
PS has also been invoked in connection with 
the striped phase of the cuprates \cite{tranq}. \\

The theoretical study of PS is a difficult subject. 
While it is a clear concept in statistical mechanics
dealing with homogeneous systems in thermodynamical equilibrium, 
the characterization
of PS in discrete systems is much more involved.  It is assumed to occur 
in those density regions where the energy as a function of density is
not a convex function. This behavior is difficult to achieve in 
finite systems where the indication of PS is a 
line,  E(x), of zero curvature, i.e of
infinite compressibility.
Even  this characterization, which should be correct if it refers to 
uniform phases of the system, is  problematic when many
inhomogeneous phases compete in the same region of parameter space. 
On the other hand, simple thermodynamic arguments suggest that
it should be a general phenomenon near magnetic phase
transitions \cite{phasesep}.\\

PS is also very hard to observe numerically as 
demonstrated by the results cited previously.
Exact results as the one obtained in \cite{nops} are very restrictive and 
hence of relative utility.\\

Our work supports the evidence for 
phase separation of the model in several ways.
The first is through the plot of the total energy
of the minimal energy configuration as a function 
of the doping $x$ shown in fig. 1. 
There we can see that the dominant feature follows
a straight line. As mentioned before, this characterization has the problem
of comparing the energies of different type of configurations. \\

The evidence is
more clear if we observe the same plot 
for a given fixed configuration in the AF region 
where phase separation occurs (fig. 2). 
The polaronic configurations in fig. 2 follow a straight line 
while negative curvature is clearly seen in the plot of the 
commensurate domain walls, the more aboundant solution in this region. 
A Maxwell construction
done to this region of curve interpolates rather well to half filling. \\

The best evidence is provided by the comparison between
the plots corresponding to the two uniform configurations
existing in the system.  In the case of the N\`eel state (AF of fig. 2)
we can  see a straight line in the 
region of densities where it  is a self-consistent solution. 
This plot should be compared with the one in fig. 10 corresponding
to  the uniform Nagaoka states.  In the large region where this 
homogeneous state is found, the plot 
follows very closely a standard quadratic curve. \\

Finally we have looked at the charge and spin 
configurations of minimal energy. Apart
from the AFM configuration at half filling and the Nagaoka FM, all inhomogeneous
configurations show the same path: coexisting regions
of an accumulation of holes  accompanied by a 
ferromagnetic order with regions of lower density with an AFM order The charge 
segregation is obvious in configurations like the ones shown in fig. 4 and 
in fig. 8c.\\

We have found fully polarized solutions in closed shell
configurations down to the lowest
electron occupancies allowed in our $12 \times 12$ cluster
(5 electrons). However, we cannot rule out the existence of
paramagnetic solutions at even lower fillings.\\

With all the previous hints we reach the conclusion that
the $t-t'$ Hubbard model tends to phase 
separate into an antiferrromagnetic and a ferromagnetic fully polarized
state with different densities for any doping away to  half
filling up to the Van Hove filling where FM sets in. 
It is interesting to note that this result was
predicted in a  totally different context by Markiewicz in ref. \cite{mark}.
Phase separation has also been predicted in the same range of dopings in ref.
\cite{ferro3} but between a paramagnetic and a ferromagnetic state.

\section{Conclusions}

In this paper we have analyzed the charge and 
spin textures of the ground state 
of the $t-t'$ Hubbard model in two dimensions as a 
function of the parameters $U$, $t'$ and
the electron density $x$ in a range from half filling to intermediate
hole doping with the aim of elucidating the role of $t'$ on 
some controversial issues. These include 
the existence and stability of ordered configurations such as domain walls
or stripes,  and the magnetic behavior in the
region of intermediate to large doping where the lower band becomes very flat.\\ 

We have used an unrestricted Hartree Fock approximation in real space
as the best suited method to study the inhomogeneous configurations of 
the system.  \\

Our results are summarized in the representative phase diagram of fig. 1
obtained for the standard values of the parameters $U=8$, $t'=0.3$. 
There we can see
that the system undergoes a  transition from 
generalized antiferromagnetic insulating
configurations including
spin polarons and domain walls, to metallic ferromagnetic configurations. 
For the values of the
parameters cited, the transition occurs at an electron density $x = 0.125$ 
($h = 18$).  
Both types of magnetic configurations converge 
in the intermediate region 
indicating that the transition is smooth more like a crossover.\\

The generalized antiferromagnetic configurations are 
characterized by a large peak in the 
density of states of
the lower band and by the presence of an  antiferromagnetic
 gap with isolated polarons
for very small doping that evolves to a mid gap subband 
for larger dopings. ferromagnetic configurations have a 
metallic character with a
DOS at the Fermi level that increases for configurations
with increasing total magnetization. Fully polarized Nagaoka states
are found at all closed shell configurations in the
ferromagnetic zone of the phase diagram. They have the 
highest DOS at the Fermi level. \\  

Apart from the homogeneous N\'eel and Nagaoka states, all inhomogeneous
configurations show the existence of the two magnetic orders associated to
charge segregation. AF is found in the regions of low charge density and
FM clusters are formed in the localized regions where the extra charge tends
to accumulate. \\

Our main conclusion is that the only stable 
homogeneous phases of the system consists
of the purely antiferromagnetic N\`eel configuration at half filling, 
and Nagaoka ferromagnetism, which appears 
around the Van Hove filling. 
We find the system is unstable towards phase 
separation for all intermediate densities. \\

We have reached this conclusion trough a careful study
of the curves representing the total energy versus doping
of the various configurations. Besides, the approach used allows us to
visualize the inhomogeneous configurations. 
In all of them we find coexisting regions
of an accumulation of holes  accompanied by a 
ferromagnetic order with regions of lower density with an 
antiferromagnetic order.\\

As the ferromagnetic phase is metallic while the N\`eel state is insulating,
we expect the transport properties of the model in the
intermediate region to resemble that of a percolating network, a system
which has attracted much attention lately \cite{SA97,SAK99}.\\

Finally, our study does not exclude the existence of other non magnetic
instabilities, most notably d-wave superconductivity. This can be,
however, a low energy phenomenon, so that the main magnetic properties
at intermediate energies or temperatures are well described
by the study presented here.

\vspace{1cm}

We thank R. Markiewicz for a critical reading of the 
manuscript with very useful comments.
Conversations held with R. Hlubina, E. Louis, and M. P. L\'opez Sancho
are  also gratefully acknowledged. This work has been supported by the 
CICYT, Spain, through grant PB96-0875 and by CAM, Madrid, Spain.

\newpage

\section{Figure captions}

Fig. 1: Complete phase diagram for U = 8, t'= 0.3. Vertical dashed lines 
separate the different configurations described in the text. Vertical 
solids lines correspond to closed shells fillings were Nagaoka 
ferromagnetism occurs. The curve is a plot of the total energy (in 
units of t) of the lowest energy configuration versus the electron 
density x.\\

Fig. 2: Comparison between the energies of the different configurations 
converging in the AF region. The configurations are displayed in fig. 4.\\

Fig. 3: Comparison between the energies of the configurations in the FM region. The configurations are displayed in fig. 8.\\
 
Fig. 4: Examples of the minimal energy configurations discussed in the text for the reference
values U=8, t'=0.3 in the antiferromagnetic region.
Fig. 4a shows the polaronic (POL) configuration obtained when doping with  five holes.
Fig. 4b shows the diagonal commensurate 
domain wall (dcDW) configuration doping with six holes. Fig. 4c shows non collinear ($S_x$)  
configuration obtained when doping with twelve holes.\\

Fig. 5: Density of states of the various configurations dicussed in the text in the 
antiferromagnetic region for the parameter values U = 8, t' = 0.3. The fermi 
level is indicated as a vertical line. Fig. 5a shows the
reference N\`eel configuration at half filling. The asymmetry of the band due to t' is
noticeable. The rest of figures  show 
POL with 5 holes (5b), POL with
14 holes(5c), dcDW with 6 holes (5d), and the $S_x$ configuration with 12 holes
(5e).\\
  
Fig. 6: Striped configurations appearing at values of the doping commensurate
with the lattice.  Fig. 6a. shows
the diagonal stripe obtained   for the parameter values $U=8$ and $t'=0.3$.
Figs. 6b and 6c correspond to a vertical stripe obtained for $U=4$ for the values
of t' $t'=0$ (6b), and  
$t'=0.2$ (6c). We can see that for bigger t' the vertical  stripe is spoiled.\\

Fig. 7: Density of states of the diagonal stripe configurations. 
The Fermi level is indicated as a vertical line.\\

Fig. 8:  Examples of the minimal energy configurations discussed 
in the text for the reference 
values U=8, t'=0.3 in the ferromagnetic region.
Fig. 8a shows the fully polarized Nagaoka state (Ng) with h=49, fig. 8b shows  
fmSDW obtained for h= 24, and fig. 8c corresponds to the  fm DOM 
obtained for h=36.\\
  
Fig. 9: Density of states of the ferromagnetic configurations in fig, 8.
Fig. 9a corresponds to the Nagaoka configuration, fig. 9b  to fmSDW, and
fig. 9c shows the DOS of FM dom. The Fermi level is indicated as a vertical line.\\

Fig. 10: Plot of the energy versus doping for the fully polarized Nagaoka configurations
in the full range of dopings where they converge. The solid line is
a fit to a quadratic curve.\\


\newpage

\begin{thebibliography}{99}

\bibitem{and} P.W. Anderson, Science {\bf 235}, 1196 (1987).

\bibitem{ttp} P. B\'enard, L. Chen and A. M. Tremblay,  Phys. Rev. B. {\bf 47}, 
15 217 (1993);
Q. Si, T. Zha, K. Levin and J. P. Lu, {\em ibid.} {\bf 47}, 9055 (1993).

\bibitem{arpes} A. Nazarenko et al., Phys. Rev. {\bf B 51}, 8676 (1995).

\bibitem{rut} Y. Maeno et al., Nature {\bf 372}, 532 (1994).

\bibitem{su} W. P. Su, Phys. Rev. {\bf B 37}, 9904 (1988).

\bibitem{pr} D. Poilblanc and T.M. Rice,  Phys. Rev. {\bf B 39}, 9749 (1989).

\bibitem{il} M. Inui and P.B. Littlewood, Phys. Rev. {\bf B 44}, 4415 (1991).

\bibitem{paco1} J. A. Verg\'es, E. Louis, P. S. Lombdahl. F. Guinea,
and A. R. Bishop, Phys. Rev. {\bf B 43}, 6099 (1991).

\bibitem{ferro1} R. Hlubina, S. Sorella, and F. Guinea,
{\it Phys. Rev. Lett.} {\bf 78}, 1343 (1997).

\bibitem{ferro3} R. Hlubina, Phys. Rev. {\bf B 59},  9600 (1999).

\bibitem{phases} J. V. Alvarez, J. Gonz\'alez, F. Guinea, 
and M. A. H. Vozmediano, 
J. Phys. Soc. Japn. {\bf 67}, 1868 (1998).

\bibitem{letal98}
E. Louis, F. Guinea, M. P. L\'opez-Sancho and J. A. Verg\'es,
Europhys. Lett. {\bf 44}, 229 (1998).

\bibitem{letal99}
E. Louis, F. Guinea, M. P. L\'opez-Sancho and J. A. Verg\'es,  
Phys. Rev. B, {\bf 59}, 14005 (1999).

\bibitem{lin} H. Q. Lin and J. E. Hirsch,  Phys. Rev. {\bf B 35}, 3359 (1987).

\bibitem{pilar} M. P. L\'opez Sancho, J. Rubio, M. C. Refolio, and J. M. 
L\'opez Sancho, Phys. Rev. {\bf B 46}, 11110 (1992).

\bibitem{capone} A. C. Cosentini, M. Capone, L. Guidoni, and G. B. Bachelet,
Phys. Rev. B {\bf 58}, R14685 (1998).

\bibitem{zaan} J. Zaanen and O. Gunnarson, Phys. Rev. B {\bf 40}, 7391 (1989).
\bibitem{Sc90}
H. J. Schulz, Phys. Rev. Lett. {\bf 64}, 1445 (1990).

\bibitem{imada} M. Imada, A. Fujimori and Y. Tokura, Rev. Mod. Phys. {\bf 70},
1039 (1998).


\bibitem{bag} J. R. Schrieffer, X. -G. Wen, and S. -C. Zhang,
Phys. Rev. Lett. {\bf 60}, 943 (1988).

 
\bibitem{vollhardt} W. Hofstetter and D. Vollhardt, 
Ann. Physik {\bf 7}, 48 (1998).

\bibitem{seib} G. Seibold, E. Sigmund, and V. Hizhnyakov, 
Phys. Rev. B {\bf 57},
6937 (1998).

\bibitem{Getal99}
F. Guinea, E. Louis, M. P. L\'opez-Sancho and J. A. Verg\'es,
{\it Universality Class of the Antiferromagnetic Transition in the
Two Dimensional Hubbard Model}, preprint (cond-mat/9901164).

\bibitem{strexperim} J. M. Tranquada, B. J. Sternlieb, J. D. Axe, Y. Nakamura
and S. Uchida, Nature {\bf 375}, 561 (1995);  J. M. Tranquada, J. D. Axe, 
N. Ichikawa, A. R. Moodenbaugh, Y. Nakamura
and S. Uchida, Phys. Rev. Lett. {\bf 78}, 338 (1997).

\bibitem{zaanen} J. Zaanen and M. Ole\'s, Ann. Physik {\bf 5}, 224 (1996).

\bibitem{seibold} G. Seibold, C. Castellani, C. Di Castro and M. Grilli, 
Phys. Rev. B {\bf 58}, 13506 (1998).

\bibitem{tt'-J} S. R. White, and D. J. Scalapino cond-mat/9812187.

\bibitem{ferro4} For a recent review with a complete list of  references see:
D. Volhardt, N. Bl\"umer, K. Held, M. Kollar, 
J. Schlipf, M. Ulmke, and J. Wahle,
Z. Phys. B {\bf 103}, 283 (1997).

\bibitem{japon} M. Murakami and H. Fukuyama,  
J. Phys. Soc. Japn. {\bf 67}, 2784 (1998);
M. Murakami, {\it Possible ordered 
states in the 2D Hubbard model}, cond-mat/9904213.

\bibitem{botton} T. Hanish, G. S. 
Uhrig, and E. M\"uller-Hartmann,  Phys. Rev. {\bf B 56},
13 960 (1997). See also 
T. Hanish, G. S. Uhrig, and E. M\"uller-Hartmann, {\em Lattice dependence of 
ferromagnetism in the Hubbard model}, cond-mat/9707286.

\bibitem{vh} A. Avella, F. Mancini, D. Villani, and H. Matsumoto,
Physika {\bf C 282-287}, 1759 (1997).

\bibitem{visscher} P. B. Visscher, Phys. Rev. {\bf B 10}, 943 (1974).

\bibitem{allps} For an overview of 
the subject, see K. A. M\"uller, and G. Benedek,
eds., Proceed. of the conf. {\it Phase separation in cuprate superconductors},
World Scientific (1993).

\bibitem{psexp} J. D. Jorgensen at al., Phys. Rev. B {\bf 38}, 11 337 (1988);
D. R. Harshman et al.,  Phys. Rev. Lett. {\bf 63}, 1187 (1989).

\bibitem{nops} G. Su, Phys. Rev. B {\bf 54}, R8281 (1996).

\bibitem{someps} V. J. Emery, S. A. Kivelson, and H. Q. Lin, Phys. Rev. Lett.
{\bf 64}, 475 (1990).

\bibitem{Cetal98}
E. W. Carlson, S. A. Kivelson, Z. Nussinov and V. J. Emery,
Phys. Rev. B {\bf 57}, 14704 (1998).

\bibitem{CBS98}
M. Calandra, F. Becca and S. Sorella, Phys. Rev. Lett. {\bf 81},
5185 (1998).

\bibitem{tj} C. S. Hellberg, and E. Manousakis, Phys. Rev. Lett.
{\bf 78}, 4609 (1997).

\bibitem{tranq} J. M. Tranquada et al., Phys. Rev. Lett. {\bf 78}, 338 (1997).

\bibitem{phasesep}
F. Guinea, G. G\'omez-Santos and D. Arovas, cond-mat/9907184 (preprint).

\bibitem{mark} R. S. Markiewicz, J. Phys. Cond. Matt. {\bf 2}, 665 (1990).

\bibitem{SA97}
E. Shimshoni and A. Auerbach, Phys. Rev. B {\bf 55}, 9817 (1997).

\bibitem{SAK99}
E. Shimshoni, A. Auerbach and A. Kapitulnik, Phys. Rev. Lett.
{\bf 80}, 3352 (1999).
\end{thebibliography}
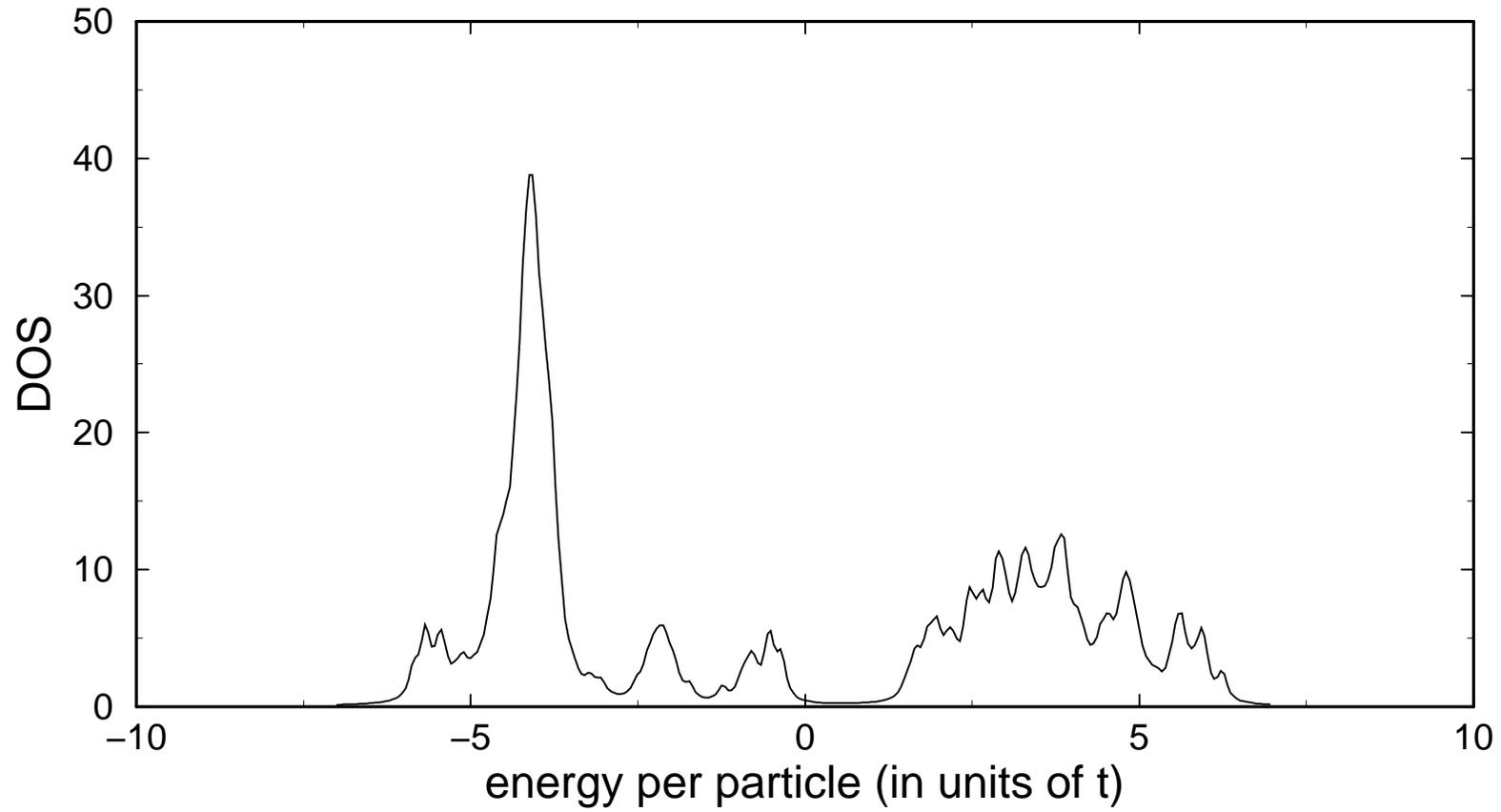
\end{document}